\documentclass[aps,prb,reprint]{revtex4-1}
\usepackage{graphicx}
\usepackage{mathrsfs}
\usepackage{bm}
\usepackage{amsmath}
\usepackage{dcolumn}
\usepackage{epstopdf}
\usepackage{dsfont}
\usepackage{amssymb}
\usepackage{tabularx}
\usepackage{array}
\usepackage{color}
\usepackage[colorlinks=true, letterpaper=true, pdfstartview=FitV, linkcolor=blue, citecolor=blue, urlcolor=blue]{hyperref}

\providecommand{\U}[1]{\protect\rule{.1in}{.1in}}

\begin{document}

\title{Chiral anomaly induced oscillations in the Josephson current in Weyl
semimetals}
\author{Salah Uddin$^2$}
\author{Wenye Duan$^3$}
\author{Jun Wang$^4$}
\author{Zhongshui Ma$^{3,5}$}
\author{Jun-Feng Liu$^{1,2}$}
\email{phjfliu@gzhu.edu.cn}

\affiliation{$^1$Department of Physics, School of Physics and Electronic Engineering, Guangzhou University, Guangzhou
510006, China}
\affiliation{$^2$Department of Physics, Southern University of Science and Technology,
Shenzhen 518055, China}
\affiliation{$^3$School of Physics, Peking
University, Beijing 100871, China}
\affiliation{$^4$Department of Physics,
Southeast University, Nanjing 210096, China}
\affiliation{$^5$Collaborative Innovation Center of Quantum Matter, Beijing,
100871, China}

\begin{abstract}
Weyl semimetals are a three dimensional topological phase of matter with
linearly dispersed Weyl points which appear in pairs and carry opposite
chirality. The separation of paired Weyl points allows charge transfer
between them in the presence of parallel electric and magnetic fields, which
is known as the chiral anomaly. In this paper, we theoretically study the
influence of the chiral anomaly induced chiral charge imbalance on the
Josephson current in a Weyl superconductor-Weyl semimetal-Weyl
superconductor junction. In Weyl superconductors, two types of pairings are
considered, namely, zero momentum BCS-like pairing and finite momentum
FFLO-like pairing. For BCS-like pairing, we find that the Josephson current
exhibits $0$-$\pi$ transitions and oscillates as a function of $\lambda_{0}
L $, where $\lambda_{0}$ is the chirality imbalance induced by the parallel
electric and magnetic fields and $L$ is the length of the Weyl semimetal.
The amplitude of the Josephson current also depends on the angle $\beta$
between the line connecting two paired Weyl points and the transport
direction along the junction. For FFLO-like pairing, the chirality imbalance
induced periodic oscillations are absent and the Josephson current is also
independent of the angle $\beta$. These findings are useful in detecting the
chiral anomaly and distinguishing the superconducting pairing mechanism of
Weyl semimetals.
\end{abstract}

\maketitle

\section{Introduction}

In the past decade or so, a great progress in condensed matter physics has
been made by the discovery of topological insulators \cite{mzhassan10,xlqi11}%
. Topological insulators have a bulk energy gap and gapless surface states,
which are protected by the time-reversal symmetry. Recently, the topological
matter is further extended to Weyl semimetals (WSMs) \cite{wanxg11}. WSM is
a three-dimensional (3D) phase where linearly dispersed Weyl cones appear in
pairs in momentum space carrying opposite chirality. The separation of Weyl
points (WPs) with opposite chirality allows charge transfer between them in
the presence of parallel electric and magnetic fields, as a consequence of
chiral anomaly \cite{sadler69,pecashby14}. The chiral anomaly is a peculiar
non-conservation of chiral charge and is mostly discussed in the context of
high-energy physics. In WSMs, the charge density at a single WP is not
conserved; the application of parallel $\mathbf{E}$ and $\mathbf{B}$ fields
drives charges from one WP to the other with opposite chirality. This charge
pumping effect induces a chemical potential difference between two paired
WPs, which is also referred to as chiral charge imbalance or chirality
imbalance. The chiral anomaly results in unusual transport properties \cite%
{phosur12,wwitczakkrempa12,ovafek14} in WSMs, which have attracted much
attention \cite%
{mmvazifeh13,rrbiswas14,yominato14,bsbierski14,ukhanna14,aaburkov15,dtson13,aaburkov14,evgorbar14,qli2016}%
.

The WSM phase requires broken time reversal \cite{kyyang11,aaburkov11} or
inversion symmetry \cite{smurakami07,amturner30}. Theoretically, WSM has
been predicted from the first principle calculation and successfully
observed experimentally in non-centrosymmetric transition metal
monophosphides, such as NbP, NbAs, TaP and TaAs \cite%
{hweng15,smhuang15,syxu15,bqlv15,syxu215,syxu315,zwang16}. Recent studies
show that WSM can also be realized in some other materials, including
pyrochlore iridates A$_2$Ir$_2$O$_7$ where A is Lanthanide or yttrium
element \cite{wanxg11}, the ferromagnetic compound HgCr$_2$Se$_4$ \cite%
{gxu11,liujy17,liuek18}, multilayer structure made of topological and
non-topological insulator thin films \cite{aaburkov11}, and magnetically
doped Bi$_2$Se$_3$ \cite{cfang12}.

The unique physics of WSMs also motivates further research on the
superconducting pairing mechanism. Particularly, doped WSMs facilitate two
types of superconducting pairings---internode and intra-node pairings. The
internode pairing forms zero momentum Bardeen-Schrieffer-Cooper (BCS) state
\cite{gycho12}, while the intra-node pairing exhibits a finite momentum
Fulde-Ferrell-Larkin-Ovchinnikov (FFLO) state \cite{gycho12,pflude64,tzhou16}%
. Different analysis methods yield different energetically preferred pairing
states. Mean-field calculations show that FFLO-like pairing is favored for
pairing states with even parity (singlet pairing) \cite{gycho12}. For
pairing states with odd parity (triplet pairing), a short and long-range
interaction results in FFLO- and BCS-like pairing states as ground states,
respectively \cite{hwei14}. The BCS-like pairing is also predicted to be
energetically preferred in the weak coupling regime \cite{gbednik15}.
Considering the theoretical controversy on the pairing mechanism in WSMs,
the experimental verification is very desirable.

The Josephson effect is one of the powerful tools to identify the
superconducting pairing. Recently, the Andreev reflection \cite{wchen13,uchida14} and Josephson
effect \cite%
{kamadsen17,ukhanna16,ukhana17,dkmukherjee17,yxu18,jfang18} in a WSM have been theoretically investigated in several works. It has been
predicted that at the interface between a time-reversal breaking WSM and a
conventional s-wave superconductor, the singlet pairing requires the
intra-node Andreev reflection due to the spin-momentum locking of Weyl
fermions \cite{ukhanna16,ukhana17,dkmukherjee17}. This extra momentum of the
pair gives rise to an unusual oscillation in the Josephson current whose
period is proportional to the distance between two paired WPs in momentum
space. The effect of quantum interference between the bulk channel and the
surface channel in Dirac semimetal based Josephson junction has also been studied
recently \cite{yxu18}. However, the investigation in the effect of chiral
anomaly on the transport in superconducting heterojunctions is still rare.

\begin{figure}[tbp]
\begin{center}
\includegraphics[bb=5 37 800 373, width=3.415in]{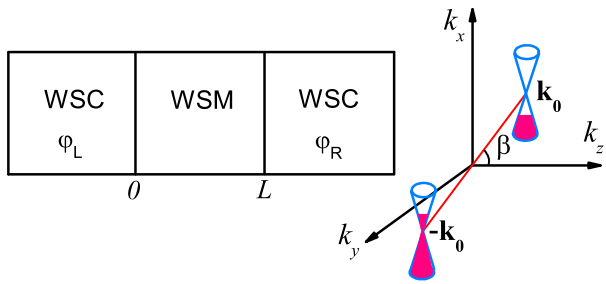}
\end{center}
\caption{(Left) Schematic diagram of the WSC-WSM-WSC Josephson junction: A
WSM at $0<z<L$ sandwiched between two WSCs at $z<0$ and $z>L$. The left
(right) WSC is characterized by the phase $\protect\varphi_L$ ($\protect%
\varphi_R)$. (Right) Schematic of the momentum space with two WPs located at
$\pm \mathbf{k}_0$ in the $k_x$-$k_z$ plane, the angle between the line
joining the two WPs and the $k_z$-axis is $\protect\beta$. There exists a
chirality imbalance between two WPs.}
\label{jjunction}
\end{figure}

In this work, we investigate the Josephson current in a Weyl superconductor
(WSC)-WSM-WSC junction as shown in Fig. \ref{jjunction}. We consider both
BCS- and FFLO-like pairings in two WSCs. For BCS-like pairing, the Josephson
current exhibits $0$-$\pi$ transitions and oscillations that depend on the
length of the WSM and the chirality imbalance induced by an $\mathbf{E}
\cdot \mathbf{B}$ field. The Josephson current also depends on the angle $%
\beta$ between the line connecting two WPs and the transport direction along
the junction. For FFLO-like pairing the $0$-$\pi$ transitions and
oscillations are absent and the Josephson current is also independent of $%
\beta$. These findings are useful in detecting the chiral anomaly and
distinguishing the superconducting pairing mechanism of Weyl semimetals.

The rest of the paper is organized as follows. In Sec. II, we introduce the
model and solve the scattering problem for quasiparticles based on the
Bogoliubov-de Gennes (BdG) equation. The Andreev bound states (ABSs) are
obtained from the scattering matrices and the Josephson current is obtained
from ABSs. In Sec. III, we present the numerical results of Josephson
current and Andreev bound states for both BSC- and FFLO-like pairing and
discuss the underlying physics. Finally, the conclusion remarks are given in
Sec. IV.

\section{Model and Formalism}

The Josephson junction under consideration is sketched in Fig. \ref%
{jjunction} with two WSCs at $z<0$ and $z>L$ and a WSM layer in the region $%
0<z<L$. The two WPs with opposite chirality are located at $\pm \mathbf{k}%
_{0}$ in the $k_{x}$-$k_{z}$ plane, and the line joining two WPs makes an
angle $\beta $ with the $k_{z}$-axis. For such a configuration, an effective
two band Hamiltonian for a single WP is used in the references \cite%
{kamadsen17,wchen13}. We use the same model for our present study. The
effective two band Hamiltonian around the Weyl node $\pm \mathbf{k}_{0}$ is
\begin{equation}
H_{\pm }=\hbar \nu (k_{1}\sigma _{1}+k_{2}\sigma _{2}\mp k_{3}\sigma
_{3})-\mu _{\pm },
\end{equation}%
where $\mu _{\pm }=\mu \mp \lambda/2$ is the chemical potential, $%
\lambda$ is the chirality imbalance induced by an $\mathbf{E}\cdot
\mathbf{B}$ field. And
\begin{eqnarray}
k_{1} &=&k_{x}\cos \beta -k_{z}\sin \beta ,  \notag \\
k_{2} &=&k_{y}, \\
k_{3} &=&k_{x}\sin \beta +k_{z}\cos \beta .  \notag
\end{eqnarray}%
Similarly
\begin{eqnarray}
\sigma _{1} &=&\sigma _{x}\cos \beta -\sigma _{z}\sin \beta ,  \notag \\
\sigma _{2} &=&\sigma _{y}, \\
\sigma _{3} &=&\sigma _{x}\sin \beta +\sigma _{z}\cos \beta ,  \notag
\end{eqnarray}%
where $\sigma _{x}$, $\sigma _{y}$ and $\sigma _{z}$ are the Pauli matrices.
The transport direction is assumed to be along the z-axis. We consider two
types of pairing mechanism \cite{gycho12}. One is the inter-node BCS-like
pairing for which two paired electrons are from two different WPs and the
Cooper pairs have zero net momentum. The other is the intra-node FFLO-like
pairing for which two paired electrons are from one single WP, and the
Cooper pairs have nonzero net momentum. In the Nambu representation with
four-component bases $[\psi _{\uparrow }(\mathbf{r}),\psi _{\downarrow }(%
\mathbf{r}),\psi _{\downarrow }^{\dagger }(\mathbf{r}),-\psi _{\uparrow
}^{\dagger }(\mathbf{r})]^{T}$, the BdG Hamiltonians for both BCS- and
FLLO-like pairing cases are given by
\begin{equation*}
H_{B}^{\pm }=\left(
\begin{array}{cc}
H_{\pm }(-i\nabla \mp \mathbf{k}_{0}) & \Delta (z) \\
\Delta (z)^{\ast } & -H_{\mp }(-i\nabla \mp \mathbf{k}_{0})%
\end{array}%
\right) ,
\end{equation*}%
\begin{equation}
H_{F}^{\pm }=\left(
\begin{array}{cc}
H_{\pm }(-i\nabla \mp \mathbf{k}_{0}) & \Delta (z)e^{\pm 2i\mathbf{k}%
_{0}\cdot \mathbf{r}} \\
\Delta (z)^{\ast }e^{\mp 2i\mathbf{k}_{0}\cdot \mathbf{r}} & -H_{\pm
}(-i\nabla \pm \mathbf{k}_{0})%
\end{array}%
\right) ,
\end{equation}%
where the subscripts $B$ and $F$ correspond to BCS- and FFLO-like pairing
respectively. And $\Delta (z)=\Delta _{0}\left[ \Theta (-z)e^{i\varphi
/2}+\Theta (z-L)e^{-i\varphi /2}\right] $ denotes the pair potential with $%
\Delta _{0}$ as the bulk superconducting gap and $\varphi =\varphi
_{L}-\varphi _{R}$ as the macroscopic phase difference between the left and
right superconductors. The temperature dependence magnitude of $%
\Delta $ is given by $\Delta (T)=\Delta (0)\tanh (1.74\sqrt{T/T_{c}-1})$ ,
where $T_{c}$ is the critical temperature. The chirality imbalance $\lambda
$ is only introduced in the normal WSM region, therefore, $\lambda(z)=\lambda _{0}\left[ \Theta (z)-\Theta (z-L)\right] $. 
The effective gap for FFLO-like pairing is just $\Delta _{F}=\Delta _{0}$, while the effective gap for BCS-like pairing is $%
\Delta _{B}=\Delta _{0}|\sin \beta |$, which depends on the angle $\beta$.

At first, we perform a gauge transformation to remove the large momentum $%
\mathbf{k}_{0}$ from the BdG equation \cite{kamadsen17}. The transformations
for BCS-like pairing and FFLO-like pairing are

\begin{eqnarray}
H_{B}^{\pm } &\longrightarrow &\widetilde{H}_{B}^{\pm }=e^{\pm i\mathbf{k}%
_{0}\cdot \mathbf{r}}H_{B}^{\pm }e^{\mp i\mathbf{k}_{0}\cdot \mathbf{r}},
\notag \\
H_{F}^{\pm } &\longrightarrow &\widetilde{H}_{F}^{\pm }=e^{\pm i\sigma _{z}%
\mathbf{k}_{0}\cdot \mathbf{r}}H_{F}^{\pm }e^{\mp i\sigma _{z}\mathbf{k}%
_{0}\cdot \mathbf{r}},
\end{eqnarray}%
respectively, which gives
\begin{eqnarray}
\widetilde{H}_{B}^{\pm } &=&\left(
\begin{array}{cc}
H_{\pm }(-i\nabla ) & \Delta (z) \\
\Delta (z)^{\ast } & -H_{\mp }(-i\nabla )%
\end{array}%
\right) ,  \notag \\
\widetilde{H}_{F}^{\pm } &=&\left(
\begin{array}{cc}
H_{\pm }(-i\nabla ) & \Delta (z) \\
\Delta (z)^{\ast } & -H_{\pm }(-i\nabla )%
\end{array}%
\right) .
\end{eqnarray}

Then, the scattering problem of such a junction can be solved by considering
the boundary conditions of the wave function at $z=0$ and $z=L$ \cite%
{jfliu10}. At each interface we get a scattering matrix, from which the
reflection matrix of the right-going (left going) incident particles $R_{1}$
($R_{2}$) can be abstracted in the WSM layer. The multiple reflections
between the WSM-WSC boundaries lead to bound state levels in the middle WSM
layer. The discrete spectrum of these ABSs along a fixed incident direction
can be determined by using the condition
\begin{equation}
\left. \text{det}[I_{2\times 2}-R_{2}PR_{1}P]\right\vert _{E=E_{b}}=0,
\label{abs}
\end{equation}%
where $P$ is the propagating matrix which accounts for the phases acquired
by the electron and hole while moving from one boundary to the other inside
the WSM region. In the short junction limit, the Josephson current is mainly
carried by the ABSs and can be estimated as
\begin{equation}
I(\varphi )=\frac{2e}{\hbar }\sum_{b}\frac{\partial E_{b}}{\partial \varphi }%
f(E_{b}),  \label{eq9}
\end{equation}%
where $f(E_{b})$ is the Fermi-Dirac distribution function. The total Josephson current can be obtained as
\begin{equation}
J(\varphi )=\frac{W^2}{(2\pi)^2}%
\int{I(\varphi)dk_x dk_y}, \label{eq9}
\end{equation}%
where $W$ is assumed to be the dimension in both $x$ and $y$ directions.

\begin{figure}[tbp]
\begin{center}
\includegraphics[bb=129 0 759 589, width=3.415in]{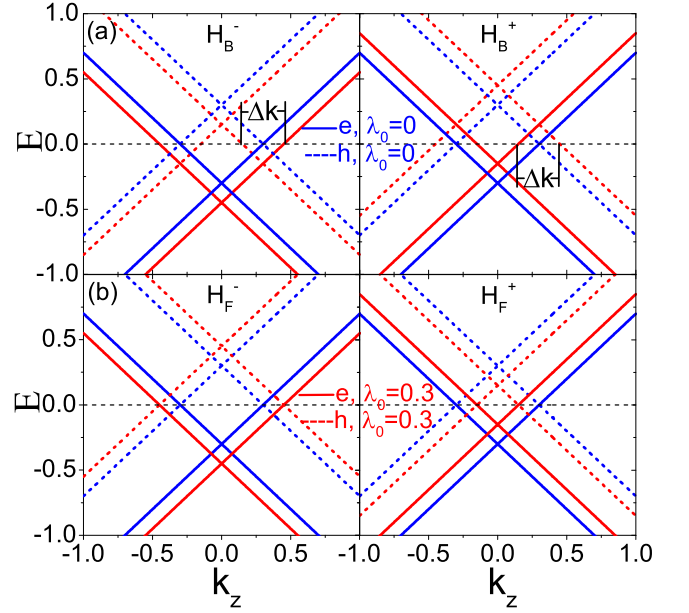}
\end{center}
\caption{Energy dispersion $E(k_z)$ of the middle WSM region: (a) for
BCS-like pairing and (b) for FFLO-like pairing with $\protect\mu=0.3$. Solid
(dashed) lines are for electrons (holes). Red (Blue) lines are for with
(without) the chirality imbalance. }
\label{band}
\end{figure}

\begin{figure}[tbp]
\begin{center}
\includegraphics[bb=129 0 759 589, width=3.415in]{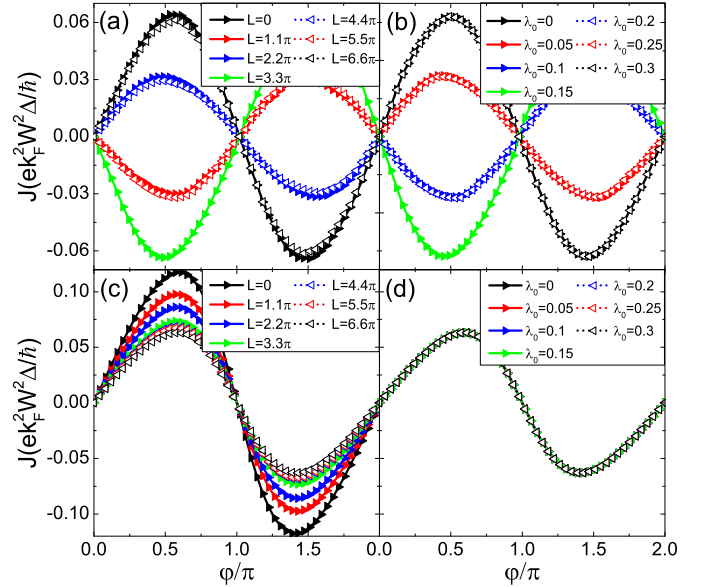}
\end{center}
\caption{Josephson current as a function of the superconducting phase
difference $\protect\varphi$. (a) and (b) are for BCS-like pairing, and (c)
and (d) are for FFLO-like pairing. (a) and (c) are for different values of
the length $L$ with fixed $\protect\lambda_0=0.3$ and (b) and (d) are for
different values of $\protect\lambda_0$ with fixed $L=6.6\protect\pi$. The
other parameters are $\protect\beta=\protect\pi/4$, $\protect\mu=0.3$ and $%
T=0.5T_c $.}
\label{jvp}
\end{figure}

\section{Results and Discussions}

Next we present the numerical results and discussion of the Josephson
current in the junction. For simplicity, we introduce the dimensionless
units: the wave vector $k\rightarrow kk_{0}$, the length $L\rightarrow
L/k_{0}$, and the energy $E\rightarrow EE_{0}$ with $E_{0}=\hbar \nu k_{0}$.
All physical quantities are expressed in these units in the rest of the
paper. The superconductors considered are characterized with $\triangle
_{0}=10^{-3}$ (in units of $E_{0}$) which correspond to the BCS coherence
length at zero temperature $\xi _{0}=2/\pi \triangle _{0}\approx 636.6$.

First, we consider the effect of $\lambda _{0}$ on the band structure of
BCS-like pairing. Fig \ref{band}(a) shows the energy dispersion of the WSM
for BCS-like pairing, with (red lines) and without (blue lines) $\lambda
_{0} $. For a given energy, the wave vectors of propagating electrons and
holes are
\begin{eqnarray}
k_{e}^{\pm } &=&\sqrt{\left[ \mu \mp \frac{\lambda _{0}}{2}+E\right]
^{2}-k_{x}^{2}-k_{y}^{2}},  \notag \\
k_{h}^{\pm } &=&\sqrt{\left[ \mu \pm \frac{\lambda _{0}}{2}-E\right]
^{2}-k_{x}^{2}-k_{y}^{2}},
\end{eqnarray}%
where $k_{e}^{\pm }$($k_{h}^{\pm }$) are wave-vectors of the right going
electrons (left going holes) for the pairs $H_{B}^{\pm }$ respectively. For
normal incidence ($k_{x}=k_{y}=0$), the wave vector difference between the
right going electrons and left going holes for the pairs $H_{B}^{\pm }$ are $%
\triangle k^{\pm }=k_{e}^{\pm }-k_{h}^{\pm }=\mp \lambda _{0}$ when $E=0$.
In the formation of Andreev bound states these difference in wave-vectors
lead to an additional phase accumulation $\triangle k^{\pm }L$ due to the
traveling of the electron and hole in the WSM layer. These additional phases
should be offset by the phase difference between the two superconductors $%
\varphi $. Therefore, the current-phase relations (CPR) for the pairs $%
H_{B}^{\pm }$ have phase shifts $\pm \varphi _{0}=\pm \lambda _{0}L$. In the
first harmonic approximation, the total Josephson current as the summation
over two pairs $H_{B}^{\pm }$ can be written as
\begin{eqnarray}
J &=&J_{H_{B}^{+}}+J_{H_{B}^{-}}  \notag \\
&=&J_{0}\sin {(\varphi +\lambda _{0}L)}+J_{0}\sin {(\varphi -\lambda _{0}L)}
\notag \\
&=&2J_{0}\sin {\varphi }\cos {(\lambda _{0}L)},  \label{jvpe}
\end{eqnarray}%
where the critical current $2J_{0}\cos {(\lambda _{0}L)}$ implies $0$-$\pi $
transitions and would change signs with increasing ${\lambda _{0}L}$. This
analytical expression of the total Josephson current gives a good fitting of
our numerical results (using Eq. (\ref{eq9})) shown in Fig. \ref{jvp}(a) and
(b). Fig. \ref{jvp}(a) shows the CPR for different length $L$ of the WSM
with fixed $\lambda _{0}=0.3$, and Fig. \ref{jvp}(b) shows the CPR for
different $\lambda _{0}$ with fixed $L=6.6\pi $. The 0-$\pi $ transitions
are clearly exhibited in Fig. \ref{jvp}(a) and (b) by increasing $L$ or $%
\lambda _{0}$.

\begin{figure}[tbp]
\begin{center}
\includegraphics[bb=129 0 759 589, width=3.415in]{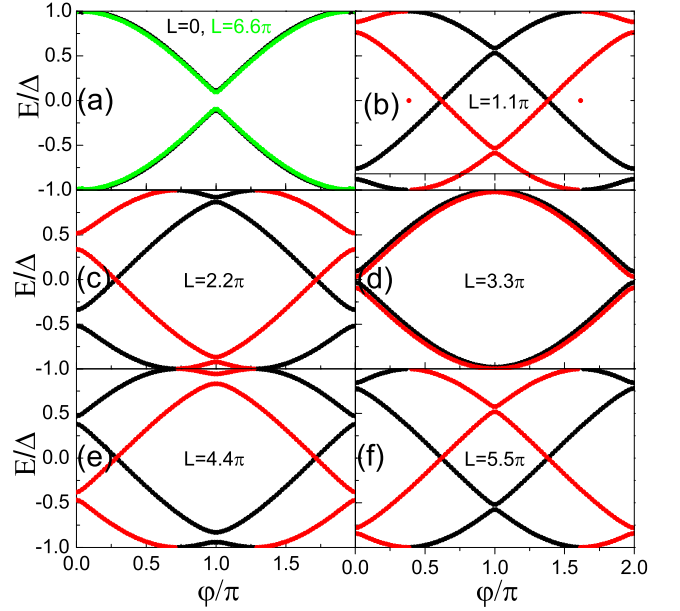}
\end{center}
\caption{Energy levels $E_b$ of the Andreev bound states (solutions of Eq. (%
\protect\ref{abs})) for BCS-like pairing. In (b)-(f), the black curves
represent bound levels for the pair $H_B^+$ and the red curves represent the
bound levels for the pair $H_B^-$. The other
parameters are the same as in Fig. \protect\ref{jvp}(a). }
\label{absf}
\end{figure}

\begin{figure}[tbp]
\begin{center}
\includegraphics[bb=129 0 759 589, width=3.415in]{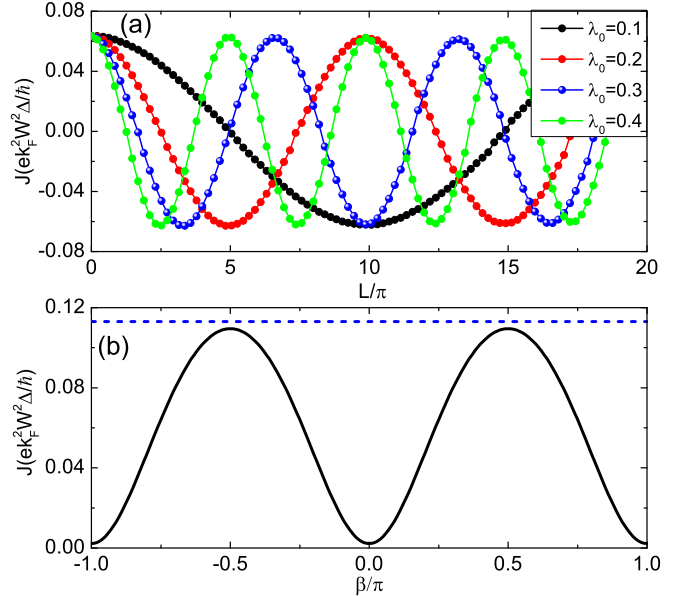}
\end{center}
\caption{(a) The Josephson current $J(\protect\pi/2)$ for BCS-like pairing
as a function of $L$ for various $\protect\lambda_0$, the other parameters
are the same as those in Fig. \protect\ref{jvp}(a). (b) The Josephson
current $J(\protect\pi/2)$ as a function of the angle $\protect\beta$, for
BCS-like pairing (black solid line) and FFLO-like pairing (blue dotted
line), the other parameters are $L=0$, $\protect\lambda_0=0.3 $, $\protect\mu%
=0.3$ and $T=0.5T_c$.}
\label{jvla}
\end{figure}

The energy dispersion of the WSM region for the case of FFLO-like pairing is
shown in Fig. \ref{band}(b). The paired electrons in FFLO-like pairing are
from the same WP, therefore, the wave-vectors difference induced by the
chirality imbalance $\lambda _{0}$ is zero (${\left\vert {\triangle k}%
\right\vert }=\left\vert \triangle k^{\pm }\right\vert =\left\vert
k_{e}^{\pm }-k_{h}^{\pm }\right\vert =0$) in each WP and the transport phase
$\varphi _{0}=0$. There is no additional phase accumulation with the
superconducting phase difference $\varphi $. In the first harmonic
approximation, the total Josephson current is
\begin{eqnarray}
J &=&J_{H_{F}^{+}}+J_{H_{F}^{-}}  \notag \\
&=&(J_{+}+J_{-})\sin {\varphi .}  \label{jpr2}
\end{eqnarray}%
The CPR for FFLO-like pairing is shown in Fig. \ref{jvp}(c) and (d). Fig. %
\ref{jvp}(c) is the CPR for different $L$ with fixed $\lambda _{0}$ and Fig. %
\ref{jvp}(d) is the CPR for different $\lambda _{0}$ with fixed $L$. The
junction is always a $0$-junction.

The chiral anomaly induced $0$-$\pi $ transitions for BCS-like pairing is
also verified by the evolution of ABSs by numerically solving Eq. (\ref{abs}%
). Fig. \ref{absf} shows the numerical results of ABSs $E_{b}$ with
increasing $L$ but fixed $\lambda _{0}$. For $L=0$, the ABSs for the two
pairs $H_{B}^{\pm }$ are degenerate as shown by the black dotted line in
Fig. \ref{absf}(a). For $L=1.1\pi $, the ABSs for the pairs $H_{B}^{\pm }$
split while the electron-hole symmetry $E\rightarrow -E$ is preserved. The
ABSs of the two pairs become degenerate again at $L=3.3\pi $ and the
junction becomes a $\pi $-junction, as shown in Fig. \ref{absf}(d). Further
increase in\ the length $L$ of the WSM splits the ABSs in opposite
directions and the degeneracy of ABSs appears again at $L=6.6\pi .$ The
junction becomes a 0-junction again as shown by the green dotted lines in Fig. \ref%
{absf}(a). This periodic evolution of the ABSs with increasing $L$ is an
indubitable evidence of $0$-$\pi $ transitions induced by the chiral anomaly.

\begin{figure}[tbp]
\begin{center}
\includegraphics[bb=129 0 759 589, width=3.415in]{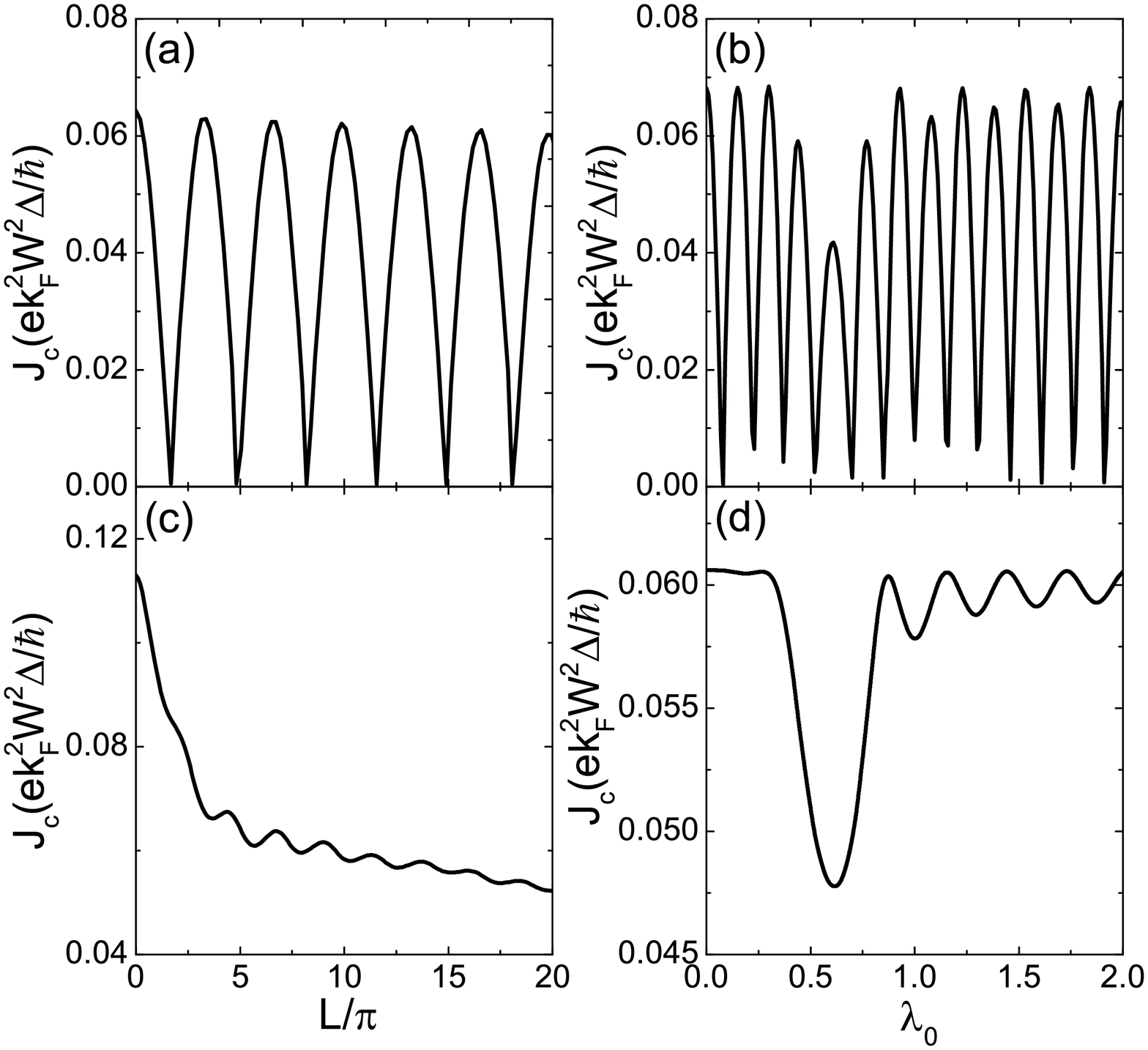}
\end{center}
\caption{(Left panel) The critical current $J_c$ as a function of the length
$L$ with fixed $\protect\lambda_0=0.3$ for BCS-like (a) and FFLO-like
pairing (c). (Right panel) The critical current $J_c$ as a function of $%
\protect\lambda_0$ with fixed length $L=6.6\protect\pi$ for BCS-like (b) and
FFLO-like pairing (d). The other parameters are $\protect\beta=\protect\pi/4$%
, $\protect\mu=0.3$ and $T=0.5T_c$.}
\label{cric}
\end{figure}

To verify the periodicity of the $0-\pi $ transitions, the Josephson current
at ${\varphi =}\pi /2$ is plotted as a function of $L$ for different values
of the chirality imbalance $\lambda _{0}$ in Fig. \ref{jvla}(a). The
periodic oscillations of the Josephson current are well fitted by Eq. (\ref%
{jvpe}). The junction is a $0$-junction for $\lambda _{0}L\approx 2n\pi $,
but a $\pi $-junction for $\lambda _{0}L\approx (2n+1)\pi $.

For the case of BCS-like pairing, the Josephson current also depends on the
angle $\beta $ between the line joining the two WPs and the transport
direction due to the effective $\beta $-dependent superconducting gap $%
\Delta _{B}=\Delta _{0}|\sin \beta |$. For $\beta =0$ and $\pm \pi $, the
WSCs have vanishing effective gap and the Josephson current would be
minimum. For $\beta =\pm \pi /2$, the effective gap is maximum and the
Josephson current would be maximum. The $\beta $-dependent Josephson current
is shown by the black curve in Fig. \ref{jvla}(b). By contrast, the
effective gap of the WSC for FFLO-like pairing is independent of $\beta $,
and the corresponding Josephson current is also independent of $\beta $ as
shown by the blue dotted line in Fig. \ref{jvla}(b).

Furthermore, we consider the easily accessible experimental signal: the
critical current $J_{c}$. Fig. \ref{cric}(a) and (c) show the critical
current as a function of the length $L$ with fixed $\lambda _{0}$ for BCS-
and FFLO-like pairing respectively. The dips in the critical current for
BCS-like pairing (Fig. \ref{cric}(a)) insure the $0$-$\pi$ transitions. By
contrast, for FFLO-like pairing, $J_{c}$ mainly decreases with increasing $L$
(Fig. \ref{cric}(c)), with small oscillations due to the normal
multi-reflection at the interfaces. A similar behavior can be found in Fig. %
\ref{cric}(b) where $J_{c}$ is plotted as a function of $\lambda _{0}$ with
fixed $L$. The constant period of the oscillations in $J_{c}$ with
increasing $\lambda _{0}$ is well consistent with $J_{c}=2J_{0}\left\vert
\cos {(\lambda _{0}L)}\right\vert $ as expressed in Eq. (\ref{jvpe}).
Therefore, the chiral anomaly induced $0-\pi $ transitions should be easily
observable in experiments by increasing $\lambda _{0}$ via increasing
parallel $\mathbf{E}$ or $\mathbf{B}$ field. On the other hand, there is a
big dip in $J_{c}$ for FFLO-like pairing (Fig. \ref{cric}(d)) which locates
at $\lambda _{0}\approx 2\mu $ where the chemical potential crosses the Weyl
point. The successive small oscillations after the big dip are also from the
transmission resonance due to the normal reflection. Therefore, this easily
accessible experimental signal, $J_{c}$, can be used to distinguish the
BCS-like pairing from the FFLO-like pairing in WSCs.

Finally, we comment on the experimental feasibility of the observation of chiral
anomaly induced supercurrent oscillations in WSMs. The unconventional
superconductivity observed in WoTe$_2$ \cite{sun15,qi16} and UPt$_3$ \cite%
{joynt02,yanase16} implies promising candidates for WSC. WSCs can also be
achieved by the proximity effect of a conventional superconductor on a WSM.
Because the model and conclusion in this work is quiet general, the
predicted chiral anomaly induced $0$-$\pi$ transition may also be observable
in such WSCs. Moreover, we suggest a thin film as the junction geometry for
the WSC/WSM/WSC junction which is narrow in the $y$ direction. The axis that
connects two WPs is suggested to be laid along the $x$ direction, because
the Josephson current is maximum at $\beta=\pi/2$ for BCS-like pairing. An $%
\mathbf{E}\cdot\mathbf{B}$ field is also applied along the $x$ axis to
induce a chirality imbalance. In such a configuration, the side effects of
the $\mathbf{E}\cdot \mathbf{B}$ field, such as the longitudinal voltage
bias of the electric field, the orbital effect of the magnetic field, and
the disturbance of the superconductivity can be well suppressed. The
separate modulation of $\mathbf{E}$ or $\mathbf{B}$ field will tune the
chirality imbalance $\lambda_0$, and therefore lead to $0$-$\pi$ transitions
and supercurrent oscillations in the case of BCS-like pairing.

\section{CONCLUSION}

In conclusion, we numerically investigate the effect of chiral anomaly on
the Josephson current in a WSC-WSM-WSC junction. We consider two types of
pairing mechanisms: the BCS-like and FFLO-like pairings. For BCS-like
pairing, the chirality imbalance $\lambda _{0}$ induces a wave-vector
difference between the electron and the hole in a Andreev bound state. This
difference in wave-vector causes a transport phase which results in $0$-$\pi $
transitions in the CPR. The critical Josephson current oscillates as a
cosine function of $\lambda _{0}L$. Furthermore, the amplitude of the
Josephson current depends on the angle $\beta $ between the line joining the
two WPs and the transport direction. However, for FFLO-like pairing, the
junction always behaves as a $0$-junction and the Josephson current is
independent of $\beta $. The experimental observation of this predicted
chiral anomaly induced $0$-$\pi $ transition in the Josephson current can
verify the chiral anomaly effect and as well as distinguish the
superconducting pairing mechanism of Weyl semimetals.

\begin{acknowledgments}
The work described in this paper is supported by the National Natural
Science Foundation of China (NSFC, Grant Nos. 11774144, and
11574045).
\end{acknowledgments}


\begin{thebibliography}{99}
\bibitem{mzhassan10} M. Z. Hasan and C. L. Kane, Rev. Mod. Phys. \textbf{82}%
, 3045 (2010).

\bibitem{xlqi11} X.-L. Qi and S.-C. Zhang, Rev. Mod. Phys. \textbf{83}, 1057
(2011).

\bibitem{wanxg11} X.Wan, A.M. Turner, A. Vishwanath, and S. Y. Savrasov,
Phys. Rev. B \textbf{83}, 205101 (2011).

\bibitem{sadler69} S. Adler, Phys. Rev. \textbf{177}, 2426 (1969).

\bibitem{pecashby14} P. E. C. Ashby and J. P. Carbotte, Phys. Rev. B \textbf{%
89}, 245121 (2014).

\bibitem{phosur12} P. Hosur, S. A. Parameswaran, and A. Vishwanath, Phys.
Rev. Lett. \textbf{108}, 046602 (2012).

\bibitem{wwitczakkrempa12} W. Witczak-Krempa and Y. B. Kim, Phys. Rev. B
\textbf{85}, 045124 (2012).

\bibitem{ovafek14} O. Vafek and A. Ishwanath, Ann. Rev. Condens. Matter Phys.
\textbf{5}, 83 (2014).

\bibitem{mmvazifeh13} M. M. Vazifeh and M. Franz, Phys. Rev. Lett. \textbf{%
111}, 027201 (2013).

\bibitem{rrbiswas14} R. R. Biswas and S. Ryu, Phys. Rev. B \textbf{89},
014205 (2014).

\bibitem{yominato14} Y. Ominato and M. Koshino, Phys. Rev. B \textbf{89},
054202 (2014).

\bibitem{bsbierski14} B. Sbierski, G. Pohl, E. J. Bergholtz, and P. W.
Brouwer, Phys. Rev. Lett. \textbf{113}, 026602 (2014).

\bibitem{ukhanna14} U. Khanna, A. Kundu, S. Pradhan, and S. Rao, Phys. Rev.
B \textbf{90}, 195430 (2014).

\bibitem{aaburkov15} A. A. Burkov, J. Phys.: Condens. Matter \textbf{27},
113201 (2015).

\bibitem{dtson13} D. T. Son and B. Z. Spivak, Phys. Rev. B \textbf{88},
104412 (2013).

\bibitem{aaburkov14} A. A. Burkov, Phys. Rev. Lett. \textbf{113}, 247203
(2014).

\bibitem{evgorbar14} E. V. Gorbar, V. A. Miransky, and I. A. Shovkovy, Phys.
Rev. B \textbf{89}, 085126 (2014).

\bibitem{qli2016} Qiang Li, Dmitri E. Kharzeev, Cheng Zhang, Yuan Huang, I. Pletikosi\'{c}, A. V. Fedorov, R. D. Zhong, J. A. Schneeloch, G. D. Gu, and T. Valla, Nat. Phys.
\textbf{12}, 550 (2016).

\bibitem{kyyang11} K.-Y. Yang, Y.-M. Lu, and Y. Ran, Phys. Rev. B \textbf{84}%
, 075129 (2011).

\bibitem{aaburkov11} A. A. Burkov and L. Balents, Phys. Rev. Lett. \textbf{%
107}, 127205 (2011).

\bibitem{smurakami07} S. Murakami, New J. Phys. \textbf{9}, 356 (2007).

\bibitem{amturner30} A. M. Turner and A. Vishwanath, arXiv:1301.0330.

\bibitem{hweng15} H. Weng, C. Fang, Z. Fang, B. A. Bernevig, and X. Dai,
Phys. Rev. X \textbf{5}, 011029 (2015).

\bibitem{smhuang15} S.-M. Huang, S.-Y. Xu, I. Belopolski, C.-C. Lee, G.
Chang, B. Wang, N. Alidoust, G. Bian, M. Neupane, C. Zhang, S. Jia, A.
Bansil, H. Lin, and M. Z. Hasan, Nat. Commun. \textbf{6}, 7373 (2015).

\bibitem{syxu15} S.-Y. Xu, I. Belopolski, N. Alidoust, M. Neupane, G. Bian,
C. Zhang, R. Sankar, G. Chang, Z. Yuan, C.-C. Lee, S.-M. Huang, H. Zheng, J.
Ma, D. S. Sanchez, B. Wang, A. Bansil, F. Chou, P. P. Shibayev, H. Lin, S.
Jia,and M. Z. Hasan, Science \textbf{349}, 613 (2015).

\bibitem{bqlv15} B.Q. Lv, H. M. Weng, B. B. Fu, X. P. Wang, H. Miao, J. Ma,
P. Richard, X. C. Huang, L. X. Zhao, G. F. Chen, Z. Fang, X. Dai, T. Qian,
and H. Ding, Phys. Rev. X \textbf{5}, 031013 (2015).

\bibitem{syxu215} S.-Y. Xu, N. Alidoust, I. Belopolski, Z. Yuan, G. Bian,
T.-R. Chang, H. Zheng, V. N. Strocov, D. S. Sanchez, G. Chang, C. Zhang, D.
Mou, Y. Wu, L. Huang, C.-C. Lee, S.-M. Huang, B. Wang, A. Bansil, H.-T.
Jeng, T. Neupert, A. Kaminski, H. Lin, S. Jia, and M. Z. Hasan, Nat. Phys.
\textbf{11}, 748 (2015).

\bibitem{syxu315} S.-Y. Xu, I. Belopolski, D. S. Sanchez, C. Zhang, G.
Chang, C. Guo, G. Bian, Z. Yuan, H. Lu, T.-R. Chang, P. P. Shibayev, M. L.
Prokopovych, N. Alidoust, H. Zheng, C.-C. Lee, S. M. Huang, R. Sankar, F.
Chou, C.-H. Hsu, H.-T. Jeng, A. Bansil, T. Neupert, V. N. Strocov, H. Lin,
S. Jia, and M. Z. Hasan, Sci. Adv. \textbf{1}, e1501092 (2015).

\bibitem{zwang16} Z. Wang, Y. Zheng, Z. Shen, Y. Zhou, X. Yang, Y. Li,
C.Feng, and Z.-A. Xu, Phys. Rev. B \textbf{93}, 121112 (2016).

\bibitem{gxu11} G. Xu, H. Weng, Z. Wang, X. Dai, and Z. Fang, Phys. Rev.
Lett. \textbf{107}, 186806 (2011).

\bibitem{liujy17} J. Y. Liu \textit{et al.}, Nat. Mater. \textbf{16}, 905
(2017).

\bibitem{liuek18} Qiunan Xu, Enke Liu, Wujun Shi, Lukas Muechler, Jacob
Gayles, Claudia Felser, and Yan Sun, Phys. Rev. B \textbf{97}, 235416 (2018).

\bibitem{cfang12} C. Fang, M. J. Gilbert, X. Dai, and B.A. Bernevig, Phys.
Rev. Lett. \textbf{108}, 266802 (2012).

\bibitem{gycho12} G. Y. Cho, J. H. Bardarson, Y.-M. Lu, and J. E. Moore,
Phys. Rev. B \textbf{86}, 214514(2012).

\bibitem{pflude64} P. Fulde and R. A. Ferrell, Phys. Rev. \textbf{135}, A550
(1964).

\bibitem{tzhou16} T. Zhou, Y. Gao, and Z.D. Wang, Phys. Rev. B \textbf{93},
094517 (2016).

\bibitem{hwei14} H. Wei, S.-P. Chao, and V. Aji, Phys. Rev. B \textbf{89},
014506 (2014).

\bibitem{gbednik15} G. Bednik, A. A. Zyuzin, and A. A. Burkov, Phys. Rev. B
\textbf{92}, 035153 (2015).

\bibitem{wchen13} W. Chen, L. Jiang, R. Shen, L. Sheng, B. G. Wang, and D.
Y. Xing, Europhys. Lett. \textbf{103}, 27006 (2013).

\bibitem{uchida14} S. Uchida, T. Habe, and Y. Asano, J. Phys. Soc. Jpn. \textbf{83}, 064711(2014).

\bibitem{kamadsen17} K. A. Madsen, E. J. Bergholtz, and P. W. Brouwer, Phys.
Rev. B \textbf{95}, 064511 (2017).

\bibitem{ukhanna16} U. Khanna, D. K. Mukherjee, A. Kundu, and S. Rao, Phys.
Rev. B \textbf{93}, 121409 (R) (2016).

\bibitem{ukhana17} U. Khanna, S. Rao, and A. Kundu, Phys. Rev. B \textbf{95}%
, 201115(R) (2017).

\bibitem{dkmukherjee17} D. K. Mukherjee, S. Rao, and A. Kundu, Phys. Rev. B
\textbf{96}, 161408(R)(2017).

\bibitem{yxu18} Y. Xu, S. Uddin, J. Wang, Z. Ma, and J.-F. Liu, Phys. Rev. B
\textbf{97}, 035427 (2018).

\bibitem{jfang18} Jun Fang, Wenye Duan, Junfeng Liu, Chao Zhang, and
Zhongshui Ma, Phys. Rev. B \textbf{97}, 165301 (2018).

\bibitem{jfliu10} J.-F Liu and K. S. Chan, Phys. Rev. B \textbf{82}, 184533
(2010).

\bibitem{sun15} Y. Sun, S.-C. Wu, M. N. Ali, C. Felser, and B. Yan, Phys.
Rev. B \textbf{92}, 161107 (2015).

\bibitem{qi16} Y. Qi, P. G. Naumov, M. N. Ali, C. R. Rajamathi, O. Barkalov,
M. Hanfland, S.-C.Wu, C. Shekhar, Y. Sun, V. S\"{u}b, M. Schmidt, E. Pippel,
P. Werner, R. Hillebrand, T. F\"{o}rster, E. Kampertt, W. Schnelle, S. Parkin,
R. J. Cava, C. Felser, B. Yan, and S. A. Medvedev, Nat. Commun. \textbf{7},
11038 (2016).

\bibitem{joynt02} R. Joynt and L. Taillefer, Rev. Mod. Phys. \textbf{74},
235 (2002).

\bibitem{yanase16} Y. Yanase, Phys. Rev. B \textbf{94}, 174502 (2016).
\end{thebibliography}
\end{document}